# Determining the effect of hot electron dissipation on molecular scattering experiments at metal surfaces

Connor L. Box[1,‡], Yaolong Zhang[2,‡], Rongrong Yin[2], Bin Jiang[2,*], and Reinhard J. Maurer[1,*].

[1]Department of Chemistry, University of Warwick, Gibbet Hill Road, Coventry, CV4 7AL, United Kingdom

[2]Hefei National Laboratory for Physical Science at the Microscale, Department of Chemical Physics, Key Laboratory of Surface and Interface Chemistry and Energy Catalysis of Anhui Higher Education Institutes, University of Science and Technology of China, Hefei, Anhui 230026, China

**ABSTRACT:** Nonadiabatic effects that arise from the concerted motion of electrons and atoms at comparable energy and time scales are omnipresent in thermal and light-driven chemistry at metal surfaces. Excited (hot) electrons can measurably affect molecule-metal reactions by contributing to state-dependent reaction probabilities. Vibrational state-to-state scattering of NO on Au(111) has been one of the most studied examples in this regard, providing a testing ground for developing various nonadiabatic theories. This system is often cited as the prime example for the failure of electronic friction theory, a very efficient model accounting for dissipative forces on metal-adsorbed molecules due to the creation of hot electrons in the metal. However, the exact failings compared to experiment and their origin from theory are not established for any system, because dynamic properties are affected by many compounding simulation errors of which the quality of nonadiabatic treatment is just one. We use a high-dimensional machine learning representation of electronic structure theory to minimize errors that arise from quantum chemistry. This allows us to perform a comprehensive quantitative analysis of the performance of nonadiabatic molecular dynamics in describing vibrational state-to-state scattering of NO on Au(111) and compare directly to adiabatic results. We find that electronic friction theory accurately predicts elastic and single-quantum energy loss, but underestimates multi-quantum energy loss and overestimates molecular trapping at high vibrational excitation. Our analysis reveals that multi-quantum energy loss can potentially be remedied within friction theory, whereas the overestimation of trapping constitutes a genuine breakdown of electronic friction theory. Addressing this overestimation for dynamic processes in catalysis and surface chemistry will likely require more sophisticated theories.

## INTRODUCTION

The Born-Oppenheimer approximation gives rise to the notion of a single potential energy surface (PES) that governs chemical dynamics. Despite its great success, the breakdown near electronic degeneracies is well known and corresponding nonadiabatic effects have profound implications in various fields such as photochemistry and single molecule electronics.[1] This is particularly true in elementary chemical reactions at metal surfaces, which are of fundamental and practical importance in heterogeneous catalysis, as there is virtually no energy threshold for electronic excitation in metals.[2] As a result, the gaseous species in the vicinity of a metal surface can easily dissipate their energy not only by exciting lattice vibrations but also through electron-hole pair excitations (EHPs).[3] Indeed, there has been growing experimental evidence of such nonadiabatic effects in surface chemistry[4] from quantum-state-resolved molecular beam scattering experiments,[5] chemicurrent measurements,[6,7] and ultrafast spectroscopy,[8] providing valuable benchmark data for testing first-principles theories of nonadiabatic gas-surface interactions.[9] However a predictive quantitation of how nonadiabatic effects contribute to measurable dynamic properties remains elusive.

The continuum of electronic states in metallic systems is a daunting challenge to the first-principles simulation of nonadiabatic gas-surface scattering dynamics.[1] While a full-dimensional quantum treatment is at present unfeasible, several pragmatic mixed quantum-classical dynamics (MQCD) methods have been developed,[10-14] two of which stand out for their practical feasibility when combined with *ab initio* electronic structure theory. The first is the independent electron surface hopping (IESH) method[10,15,16] which is based on the popular surface hopping trajectory method[17] that characterizes nonadiabatic effects via probabilistic electronic transitions between electronic states.[18] The IESH method describes the hopping of independent electrons with a Newns-Anderson Hamiltonian parametrized with density functional theory (DFT) data. However, it is difficult to determine excited states and their nonadiabatic couplings from first principles for metallic systems and several *ad hoc* approximations are required in the parametrization.[15] An alternative is the molecular dynamics (MD) with electronic friction (MDEF) method which assumes weak nonadiabaticity[19,20] (Eq. 1). Herein, electronic degrees of freedom (DOFs) are described via a frictional damping force that represents the nonadiabatic linear response of electrons to the motion of adsorbate atoms.[21] This force acts on the atoms in addition to the force arising from the PES (Eq. 1).

$$M\ddot{R}_i = -\frac{\partial V(\boldsymbol{R})}{\partial R_i} - \sum_j \Lambda_{ij}\dot{R}_j + \mathcal{R}_i(t) \qquad (1)$$

In (Eq. 1), $M$ and $R$ are the mass and position of a nucleus, respectively, $i$ and $j$ are nuclear coordinates, $V$ is the PES, $\Lambda$ is the electronic friction tensor (EFT) and $\mathcal{R}$ is a force associated with random white noise from the bath of electrons. In practice, the MDEF method is always further approximated by imposing the Markov approximation of instantaneous response in the constant coupling limit,[20] where some pragmatic assumptions are made in how the friction tensor is calculated from DFT in that limit.[12] Examples of approximations include the local density friction approximation (LDFA), which allows for an efficient calculation of scalar isotropic friction from the electron density of the metal[22, 24-30] and the more realistic orbital-dependent friction (ODF),[12, 23] which is calculated from Kohn-Sham DFT via time-dependent perturbation theory to provide a better description of the mode-selective nature of nonadiabatic molecule-metal energy transfer. [24-28]

All existing practical methods to study nonadiabatic dynamics introduce significant approximations which need to be scrutinized against experiment. This is of course mixed with the underlying errors of adiabatic PES itself in describing the energy landscape. Unfortunately, there is little quantitative data for realistic systems that describes under which conditions exactly which approximation breaks down and how this depends on the molecule-metal coupling strength. In other words, how do we know when the weak-coupling limit is satisfied and when the MDEF method is reliable for a particular system? While this question was partially addressed by Dou and Subotnik for simple model systems,[29] here we provide quantitative insights for state-to-state scattering of NO from Au(111), which has been considered a representative strong-coupling showcase for the breakdown of electronic friction theory.[30, 31]

Over many years, Wodtke and coworkers have collected ample state-to-state experimental data that reveals unambiguous nonadiabatic characteristics of this benchmark gas-surface process for a wide range of scattering conditions,[5, 9, 31-36] stimulating many different theoretical studies.[10, 15, 37, 38] While the aforementioned IESH model has partially accounted for the multi-quantum vibrational relaxation/excitation of NO scattered from Au(111),[16, 33] its predictions on the translational energy dependence of vibrational relaxation probabilities[36] and some other observables[31] were qualitatively inconsistent with experimental findings. These discrepancies have been largely attributed to the "too-soft" and "too-corrugated" adiabatic PES within the diabatic model Hamiltonian expressed by simple pairwise potentials.[36] Using the same adiabatic PES, earlier MDEF calculations have qualitatively failed to describe the nonadiabatic dynamics for this system, especially the vibrational excitation of NO($v_i$=0) scattering from Au(111) in the vibrational ground state.[33] Interestingly, a reduced-dimensional quantum-mechanical version of electronic friction has been able to yield broad vibrational state distributions compatible with experiment.[38]

The recent emergence of high-dimensional machine-learning-based PES has enabled the reduction of interpolation errors stemming from the fitting of the PES.[39] In this work, we use a recently developed embedded atom neural network (EANN) based adiabatic PES of this system with high-fidelity involving realistic surface DOFs.[40] With a more accurate description of repulsive NO-Au(111) interaction and potential energy topography shaped by the tight transition state, this PES has enabled much more adiabatic vibrational energy transfer than previously expected and provides a qualitatively correct translational energy dependence of vibrational inelasticity.[40] This allows us to largely reduce errors in the description of the adiabatic PES and to focus on scrutinizing the quality of nonadiabatic description at a level that was not possible before. By combining this highly accurate PES with a faithful multi-dimensional EANN representation of the full-rank ODF EFT derived by time-dependent perturbation theory,[41] in the present work, we study systematically the influence of EF on the state-to-state scattering dynamics of NO from Au(111) (see Figure 1 for a schematic system definition). Impressively, incorporating both molecular and surface DOFs, the MDEF(ODF) model allows a quantitatively correct description of the single quantum vibrational relaxation dynamics of NO($v_i$=3) and NO($v_i$=2) and their dependence on translational energy. The MDEF(LDFA) model, by contrast, has little effect on the dynamics beyond the adiabatic description. We provide a detailed analysis to rationalize this. By comparing against experimental data with systematically increasing incidence vibrational energy, we further pinpoint the energetic regime in which MDEF and, specifically Markovian MDEF, break down.

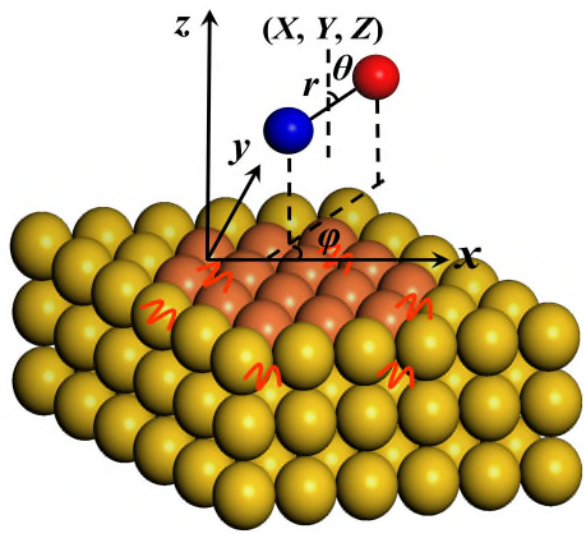

Figure 1. A schematic plot of NO on Au(111) showing internal ($X$, $Y$, $Z$, $r$, $\theta$, $\varphi$) coordinates of NO molecules and relaxation of surface atom.

## RESULTS & DISCUSSION

**General performance of electronic friction methods.** State-to-state quantum scattering of NO from Au(111) is a perfect system to scrutinize the performance of MDEF due to the availability of experimental data for various different incidence conditions. Figure 2 shows final vibrational state ($v_f$) distributions for NO scattering prepared in initial vibrational states ($v_i$) of 2, 3, 11, and 16.[31, 32, 35, 36] All experiments were performed at incidence energies ranging between 0.52 and 1.08 eV. Initial rotation states are chosen to closely match that employed in experiment. The scattering events are expected to be dominated by a single bounce due to the narrow angular distribution observed in experiment.[5, 36] We first compare experiments with IESH and MDEF simulations performed on a previously pub-lished PES.[31] The IESH simulations, show strong overestimation of the elastic scattering contributions for $v_i$=11 and $v_i$=16. MDEF simulations on the same PES correctly predict the elastic scattering populations but deliver vibrational state distributions that only lose 2-4 quanta on average with almost no population at lower final vibrational states. The failure of both methods is evidence of an inaccurate PES.[44]

Figure 2 further shows the results of adiabatic scattering simulations (labelled Born-Oppenheimer MD, BOMD), and MDEF simulations with LDFA and ODF using the new high-dimensional EANN PES and representation of EFT (the construction of which is described in the method section and the SI). We find that the angular scattering distribution predicted by the EANN model is in good agreement with the experiment, with a low population of multibounce events (see Figure S6). Recent theoretical studies with other less accurate PESs[36, 42] showed the necessity of excluding multibounce trajectories to acquire more realistic results. It is not necessary to exclude multibounce trajectories with our EANN PES, we do so anyway for the vibrational state distributions shown in the main text to ensure that vibrational energy loss only arises due to scattering and not due to equilibration on the surface. For completeness, Figures 2-5 are reproduced in the SI with multibounce events included. (Figures S7-S10) As expected, the distributions do not differ significantly. We further exclude from our analysis any vibrational states whose populations were not measured in the corresponding experimental work[43] (see Figure S7-S10 for an analysis featuring all simulated final states).

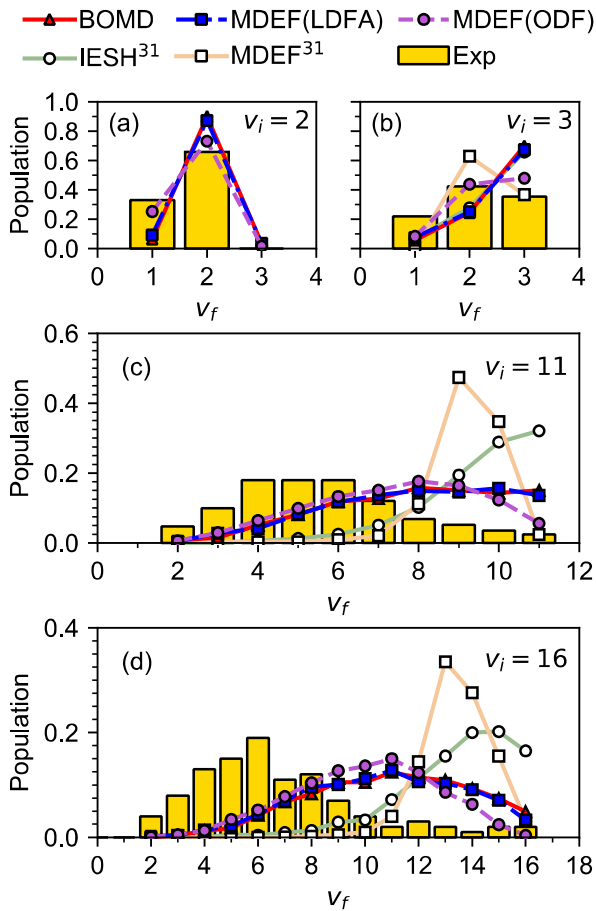

Figure 2. Experimental (Exp, golden histogram bars, respective references) vs BOMD vs LDFA vs ODF final vibrational state distributions for (a) $v_i$=2, $j_i$=2, $E_i$=0.640 eV,[32] (b) $v_i$=3, $j_i$=0, $E_i$=1.08 eV,[36] (c) $v_i$=11, $j_i$=0, $E_i$=0.950 eV,[35] and (d) $v_i$=16, $j_i$=0, $E_i$=0.520 eV.[35] Only single bounce trajectories are included in all models including the reference IESH and MDEF data,[31] additionally all data has been renormalized to the experimental limits (See Figure S7 for further clarification). The referenced MDEF friction coefficients have been calculated with a different approximate approach that is not comparable to ODF or LDFA. Lines are drawn between markers for visual clarity.

Our results correctly predict that NO scattering is highly vibrationally inelastic featuring the loss of one or more vibrational quanta leading to very broad final state distributions of highly vibrationally excited NO molecules that are clearly dominated by multi-quantum vibrational energy loss. Our tensorial orbital-based description of EF is a significant improvement over the adiabatic description and local-density description of EF for single vibrational quantum loss and elastic scattering for low initial vibrational states. Impressively, in all conditions, the new MDEF(ODF) results agree better with experiment than the reference IESH and MDEF data in terms of the broadness/shape and peak positions of final state distributions, despite some remaining discrepancies with experiment. This implies that previous studies on the failure of EF theory might have conflated PES artefacts with failings to describe nonadiabatic effects. Indeed, the current more accurate PES allows us to isolate the role of nonadiabatic effects by analyzing the remaining discrepancies of MDEF(ODF) simulations with experiment. We identify two major discrepancies, namely i) MDEF(ODF) fails to improve on the adiabatic description and continues to underestimate multi-quantum vibrational energy loss for low and high incidence vibrational energies (e.g the underestimation of $v_f$=1 population for $v_i$=3 shown in Figure. 2b) and ii) MDEF(ODF) overestimates the trapping probability (see Figure 6). The failure to reproduce the $v_f$=1 population for $v_i$=3 or make any significant improvement over the adiabatic description is particularly telling for the inability of MDEF to predict multi-quantum vibrational energy loss, which we analyze in detail further below. This is further emphasized in Figure 2b-c where MDEF(ODF) only really modifies the populations of elastic and single quantum loss channels. We also note that the MDEF(ODF) description slightly overestimates the elastic population for $v_i$=11 but underestimates it for $v_i$=16. The adiabatic results (previously discussed by some of us[40]) capture a significant portion of multi-quantum loss for $v_i$=11 and $v_i$=16, where the dominant vibrational scattering channels are 3 and 5 quanta loss respectively, though this is around half of what is predicted by experiment.

The behavior exhibited in Figure 2a-b for low initial vibrational states largely holds for a range of incidence translational energies as demonstrated in Figure 3. MDEF(ODF) performs well across the range of incidence energies for elastic and single-quantum inelastic scattering, capturing best the translational energy dependence of vibrational inelasticity among all theoretical models. However, it only slightly improves upon the adiabatic description of two-quantum inelastic scattering, with both only managing to capture the general qualitative trend that high incidence energies lead to more two-quantum loss. Interestingly, MDEF(ODF) performs slightly worse at lower incidence energies (Figure 3a,c). We can attribute this to an artifact of the underlying PES (Figure S8) which will be discussed below. MDEF(LDFA) fails to significantly improve upon the adiabatic description at any incidence energy, strongly suggesting an account of the molecular nature of the impinging NO and the directional dependence and inter-mode coupling of EF is required to describe this system. The adiabatic results capture the qualitative incidence energy dependence, though significantly overestimate the importance of the elastic channel. In the following, we will investigate the origin of the failures of our EF simulations.

**Failure to predict multi-quantum loss for low initial vibrational states.** We can further analyze the origin of the failure to capture multi-quantum loss by breaking down the $v_i=3$, $E_i=0.950$ eV experiment with respect to initial molecular orientation as the experiment shows a significant orientation dependence. In the experiments, the NO molecules are aligned with the nitrogen (N-down) or with the oxygen (O-down) pointing towards the surface. Figure 4 shows that molecules that start with an N-down orientation experience significantly more inelastic energy loss than molecules that start with O-down. MDEF(ODF) simulations succeed in reproducing this effect qualitatively, whereas adiabatic and MDEF(LDFA) models do not.

A closer look reveals that the final state distribution of O-down scattering is particularly well reproduced by MDEF(ODF), albeit with some level of underestimation of multi-quantum loss. Experiments reveal that N-down dynamics undergo more single and double vibrational quanta loss; MDEF(ODF) reproduces the former well but not the latter whilst overestimating the elastic contribution. It appears that the inability to describe sufficient multi-quantum loss from N-down dynamics is the major source of discrepancy between MDEF(ODF) and experiment for $v_i=3$ scattering shown in Figures 2 and 3.

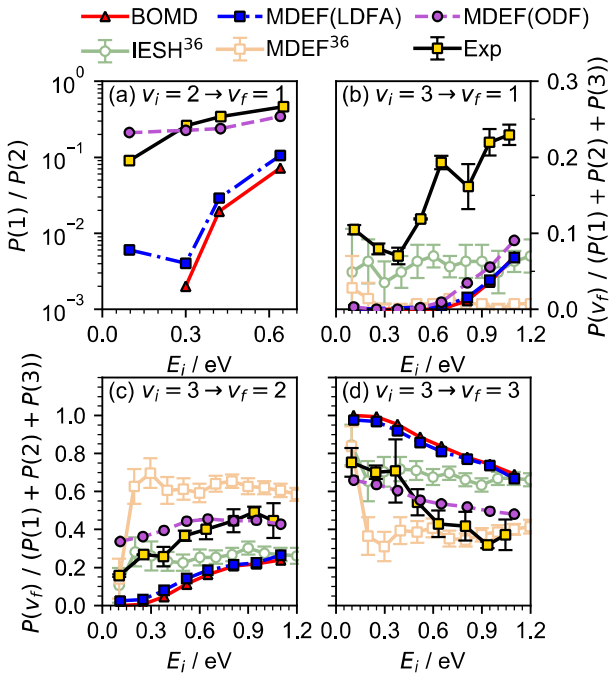

Figure 3. Experimental (Exp, relevant references) vs BOMD vs LDFA vs ODF branching ratios (for population of final state, $P(v_f)$) for $v_i=2$ ($j_i=2$)[32] and $v_i=3$ ($j_i=0$).[36] Each plot is labelled with an arrow from the initial state to the final state. Only single bounce trajectories are included in all models including the reference IESH

and MDEF data.[36] BOMD predicts no $v_i=2$ to $v_f=1$ population at low incidence energies so is omitted. Lines are drawn between markers for visual clarity.

We note that, in agreement with other theoretical results[16] for this system, there is a strong dynamical steering effect, such that in our results an initial orientation does not guarantee a similar orientation when colliding with the surface. N-down collision geometries are energetically preferred, such that even O-down initially orientated trajectories are predominantly steered into an N-down collision geometry. (Figure S11a) On average, we can see initial N-down orientations correspond to closer approaches to the surface and higher elongation of the N-O bond. (Figure S11b) Already an EF model as simple as LDFA tells us that nonadiabatic effects increase exponentially as molecules come closer to the surface and this leads to stronger nonadiabatic molecule-metal coupling. From our previous work on $H_2$ on Ag(111), we know that bond elongation leads to drastic increases in nonadiabatic coupling along the intramolecular stretch mode.[24] We further see that trapping predominantly occurs for trajectories with a close surface approach of <2 Å. (Figure S12) We reasonably expect the trapping behavior for $v_i=3$ to be similar to that experimentally determined for $v_i=2$, which is expected to be negligible for $E_i=0.950$ eV.[44]

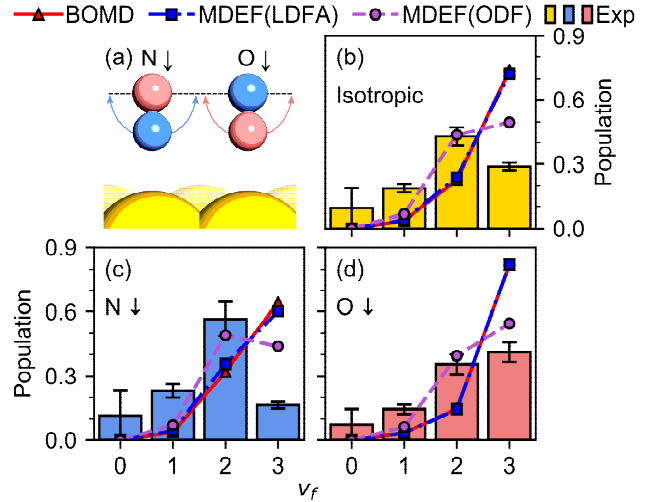

Figure 4. (a) Depiction of N (blue) down and O (red) down orientations. Experimental vs BOMD vs LDFA vs ODF final vibrational state distributions for $v_i=3$, $j_i=0$, $E_i=0.950$ eV[45] with (b) all, (c) only nitrogen down and (d) only oxygen down orientations included. Only single bounce trajectories are included. Lines are drawn between predicted distributions for visual clarity. Experimental results are shown as histogram bars. Note that in the experimental work, the $v_f=0$ population is not explicitly measured but rather assumed to be one half of the corresponding $v_f=1$ population in all cases.

At close surface approach, several effects could contribute to the underestimation of multi-quantum loss: first, molecules could be trapped that should in fact scatter with substantial energy loss. We discuss this effect further below. A second effect could lie in the current calculation of the ODF EFTs which only considers excitations that are both i) first-order (single-electron excitation) and ii) interband (ie. transitions that conserve momentum). It has been shown that including phonon-assisted intraband excitations are the dominant contributor to the short vibrational lifetime of CO adsorbed on a Cu(100) surface[46], though at a dense coverage of adsorbate molecules which is not the case here. A possible neglect of intraband contributions will particularly affect lower vibrational states and lower translational energies. We can test this effect by increasing the size of the unit cell, which effectively increases the number of electron-hole-pair excitations in the Brillouin zone that are accessible by momentum-conserving excitations. The effect is explored in detail in Figure S4. When changing unit cell size, the EFT elements do not change drastically over a range of energies when no broadening is used nor does the broadened EFT significantly change. This suggests that intra-band contributions are sufficiently accounted for in our description.

Lastly, practical MDEF simulations are always performed within the Markov approximation. The time-dependent motion of the adsorbate excites EHPs and the ensuing energy dissipation between these DOFs is dependent on the energy of the perturbing molecular motion and the density of states (DOS) of the substrate. Due to the Markov approximation, here we assume that it is independent of both. In the following, we will explore how this affects our results for high initial vibrational states.

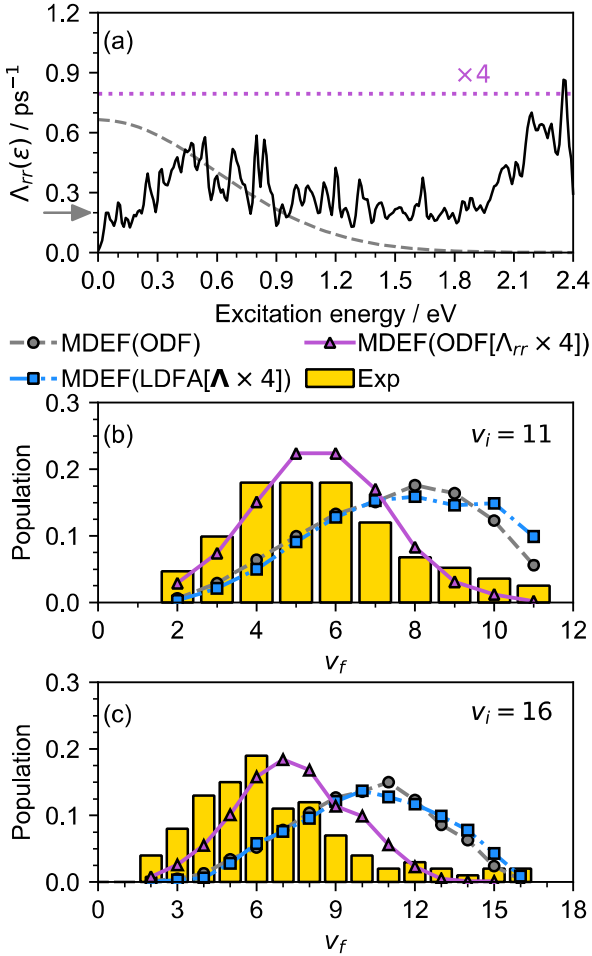

Figure 5. (a) Mass-weighted internal stretch friction excitation spectrum for the adsorption structure defined in the SI. The gray dashed line is a 0.6 eV Gaussian curve used when evaluating the internal stretch element value (gray arrow), whilst the horizontal purple dotted line depicts the element multiplied by 4 as employed when scaling ODF. ODF with and without an anisotropic scaling of 4 vs LDFA with an isotropic scaling of 4 are shown for (b) $v_i$=11

and (c) $v_i$=16,[35] conditions are otherwise the same as Figure 2. Single bounce selected only. Lines are drawn between markers for visual clarity.

**Failure to predict multi-quantum loss for high initial vibrational states.** In the case of highly vibrationally excited molecules ($v_i$=11 and $v_i$=16), the failure of MDEF(ODF) to predict multi-quantum loss upon scattering is even more evident (see Figure 2c-d). A decreased sensitivity of the final vibrational state distribution to both incidence energy and molecular orientation was observed in experiment for $v_i$=11 and further for $v_i$=16.[35] This was suggested to be due to the driving force of vibrational relaxation becoming very large due to the high vibrational state.[35]

To understand how the failure of MDEF(ODF) occurs, we must recall how the EFT is calculated. The EFT element $\Lambda_{ij}$ associated with adsorbate motion in directions $R_i$ and $R_j$ in first order perturbation theory can be expressed as,[12, 19]

$$\Lambda_{ij}(\hbar\omega) = 2\pi\hbar \sum_{k,\nu,\nu'>\nu} \left\langle\psi_{k\nu}\left|\frac{\partial}{\partial R_i}\right|\psi_{k\nu'}\right\rangle\left\langle\psi_{k\nu'}\left|\frac{\partial}{\partial R_j}\right|\psi_{k\nu}\right\rangle \\ \cdot \left(f(\varepsilon_{k\nu}) - f(\varepsilon_{k\nu'})\right) \\ \cdot (\varepsilon_{k\nu'} - \varepsilon_{k\nu}) \cdot \delta(\varepsilon_{k\nu'} - \varepsilon_{k\nu} - \hbar\omega) \quad (2)$$

The EFT for adsorbate motion with a frequency associated with energy $\epsilon = \hbar\omega$ is calculated by summing over the product of relevant nonadiabatic coupling matrix elements over all possible excitations between effective single particle Kohn-Sham states $\varepsilon_{k\nu}$ and $\varepsilon_{k\nu'}$ with respective occupation factors $f(\varepsilon_{k\nu})$ and $f(\varepsilon_{k\nu'})$. By assuming a constant DOS around the Fermi level, Head-Gordon and Tully,[20] were able to invoke a constant coupling assumption, which leads to the Markovian expression of MDEF. In practice, the EFT is evaluated at the Fermi level (zero excitation energy), replacing the delta function with a smearing function of 0.6 eV finite width.[12] Lifting the Markov approximation would lead to the inclusion of memory effects, which corresponds to the inclusion of EF at higher perturbing energies due to the modulation of particle velocity during the scattering trajectory. The inclusion of memory effects in the electronic friction force leads to a response between EHPs and adsorbate DOFs that draws contributions from the full EF spectrum. We expect that the importance of EF at higher perturbing energies will increase for higher incidence vibrational energies.

The friction excitation spectrum (Figure 5a) shows that for small broadening values and perturbing energies other than zero, the coupling may reach values several times higher than the Markovian EFT value indicated by the arrow. High vibrational states of NO lead to strong velocity oscillations and the excitation of EHPs further away from the Fermi level. As can be seen in the spectrum, the constant coupling approximation is not a good one in the case of NO on Au(111). No full memory-dependent implementation of MDEF exists at the moment, but it is clear that the inclusion of memory effects will lead to an increase in the magnitude of electronic friction and the Markovian EFT corresponds to a lower bound. We can investigate the potential effects of memory by scaling the ODF EFT internal stretch element by 4 (see SI for methodology), which approximatively represents the difference between the broadened Markovian friction value and the highest friction values present in the spectrum at non-zero frequencies. In this manner, we are studying close to an upper bound of the effects of memory on the strength of EF forces. Further in the SI, we demonstrate that the internal stretch element governs the nonadiabatic vibrational distribution with very little difference between an isotropic scaling of the whole ODF EFT or just the internal stretch element. (Figure S10)

Though the individual state populations described by the scaled MDEF(ODF) model for $v_i$=11 and $v_i$=16, presented in Figure 5b-c, show deviations in relative contribution from experiment of about 0.05-0.10, the overall vibrational distribution is well represented. A similar scaling for the $v_i$=3 case also shows an improvement of the final state distribution (see Figure S9). Notably, scaling of the LDFA EFT does not provide any improvement on the results presented in Figure 2, which again confirms that the anisotropic nature of EF must be accounted for. Scaling the EFT is of course a primitive approach to account for the nonadiabatic coupling that arises from the excitation of EHPs at various energies present within this system, we instead use it for qualitative analysis of the shortcomings of MDEF(ODF) and its comparison to MDEF(LDFA). A more advanced representation of dynamical energy loss by including the memory-dependence of EF would likely provide a more accurate final state distribution. If confirmed, this would mean that the energy loss of gas-surface scattering, even when it involves high vibrational excitation, can be represented without having to abandon the conceptual basis of electronic friction theory. This stands in contrast to previous experimental and theoretical works,[30] which assign the inability to correctly represent the final state distribution to a direct metal-molecule electron transfer due to the presence of a transient anionic state of NO. This would, however, correspond to a clear departure from the weak coupling limit described by MDEF towards multistate dynamics as described by trajectory surface hopping techniques such as IESH.

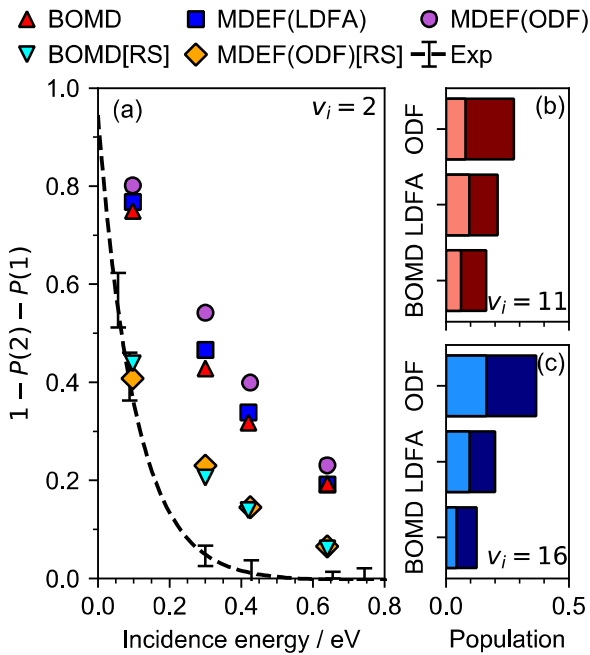

Figure 6. (a) Experimentally determined and BOMD, LDFA and ODF predicted trapping probabilities for $v_i$=2 ($j_i$=2) over a range of incidence energies.[44] Also shown are BOMD and ODF with a rescaled potential surface, [RS]. The black dashed line represents an experimentally determined fit.[44] The absolute trapped populations for (b) $v_i$=11, $j_i$=0, $E_i$=0.950 eV and (c) $v_i$=16, $j_i$=0, $E_i$=0.520 eV are also shown. Lighter bars are recorded with the rescaled potential energy surface.

**Failure to capture the trapping probability.** Finally, we wish to discuss the failure of MDEF to describe the trapping probability of NO scattering from Au(111). Trapping probabilities are overestimated from those that were experimentally determined for $v_i$=2, assuming the $v_f$=3 (excitation) and $v_f$=0 (double quantum loss) channels are negligible, the former has been experimentally recorded to be very small.[32] We employ the same methodology to calculate the model predicted trapping probabilities in Figure 6a. Indeed, the predicted $v_f$=0 and $v_f$=3 populations are very small (see Figure S7a) so that the trapping probability is very close to the absolute trapped population. Figure 6a shows the systematic overestimation of trapping over a range of incidence energies for adiabatic dynamics, with the application of either friction model not changing the picture significantly. The two possible origins of this are a potential overestimation of the adsorption well in the EANN PES rooted in the intrinsic errors of the semi-local PW91 functional or the presence of strong nonadiabatic effects such as transient ion formation that leads to a dynamical change in the energy landscape.[16, 47] We expect the former to affect low incidence energy scattering more strongly and the latter to affect high incidence energy scattering more strongly.

Indeed, we find that the EANN PES for NO on Au(111), while substantially more accurate than previous models, does still overestimate the molecule-surface attraction. In order to identify if the overestimation of trapping is due to the energy landscape or due to the description of nonadiabatic effects, we have added a repulsive contribution to the PES to reduce the adsorption energy to the experimentally observed value of 0.24 eV[48] (see SI for details). This results in a trapping probability at low incidence energies that is very close to the experimentally observed value (see ODF[RS] in Figure. 6). In the SI, we show that the adjusted PES only has minor effects on final vibrational state distributions, with the exception of improving the agreement of MDEF(ODF) with experiment for low incidence energies (see Figure S8). Figures S9-10 show that the final state distributions for $v_i$=3, 11 and 16 at moderate incidence energies originally shown in Figure 2 and 4 are not significantly affected, leaving our previous conclusions on multi-quantum energy loss unaffected. This also suggests that our main conclusion in this work would not be significantly altered using different density functionals that may yield different adsorption well depts or barrier heights. Nevertheless, as can be seen in Figure 6a, the trapping probability as predicted by MDEF(ODF[RS]) remains too high at high incidence energies.

Figure 6b-c show the absolute trapped population for both $v_i$=11 and $v_i$=16. The trapping probability has not been measured experimentally for high incidence vibrational states, though it is expected to be relatively insignificant, even at very low incidence energies ($E_i$=0.05 eV).[5, 49] On this basis, considering the high vibrational state and the moderate to high incidence energies employed, we should expect negligible trapped population. This is not the case for the adiabatic results and the application of EF which leads to an even larger trapping probability. Application of the rescaled PES significantly reduces the trapping (Figure 6b-c) but does not nullify it.

Contrary to our results, the low trapping probability at high $v_i$ has been correctly predicted by Shenvi et al.,[16] where IESH predicts a trapping probability far lower than their BOMD results and far lower than what our present BOMD and MDEF results suggest. The lowering of the trapping probability compared to adiabatic results has been related to transient nonadiabatic metal-to-molecule charge transfer which leads to an enhancement of vibration-to-translation energy transfer.[16] This is opposite to the effect that electronic friction has, which dominantly describes vibrational dissipation into EHPs, enhancing molecular trapping rather than reducing it.

## CONCLUSION

We have presented a systematic analysis of the performance of state-of-the-art nonadiabatic simulation methods in describing hot-electron effects in vibrational state-to-state-scattering of NO on Au(111). To understand how nonadiabatic effects contribute to measurable dynamic reaction probabilities, we need to be able to isolate the role of nonadiabatic effects from other contributing factors. This is made possible with a newly created high-dimensional machine-learning-based potential energy landscape that resolves artefacts of previous PES models. While the model still overestimates the probability of trapping, readjustment of the PES to match experimental trapping at low incidence energies shows that the quantities of interest, namely final vibrational state distributions upon scattering, are not strongly affected by this. Using a rotationally covariant machine-learning model, we construct a high-dimensional model of ODF electronic friction calculated from DFT. We find that MDEF(ODF) provides excellent agreement with experiment for elastic scattering of various initial vibrational states and for single quantum vibrational energy loss of low initial vibrational states ($v_i$=2 and $v_i$=3), as well as the orientation dependence of vibrational state distribution, but otherwise underestimates multi-quanta vibrational energy loss. Particularly in the case of high initial vibrational states such as $v_i$=11 and $v_i$=16, the width and the shape of the final state distribution is well described, but the average number of lost vibrational quanta is much smaller than in experiment. As we apply the Markov approximation, the high-lying EHPs of 1.5 eV and beyond that are excited by such high vibrational states are not included in our EF description. By analysis of the friction spectrum and rescaling of the friction tensor to account for this shortcoming, we find that we can reproduce the overall population distribution of inelastic scattering, albeit at a small remaining underestimation of the proportion of scattering outcomes with low vibrational energy. This finding is surprising as it suggests that a full account of memory effects within EF theory could potentially extend the remit of MDEF to describe the final vibrational state distributions of highly excited molecules without having to resort to the explicit inclusion of strong nonadiabatic effects on the energy landscape.

However, memory-dependent MDEF would likely not resolve the second failure of our MDEF results, namely the significant overestimation of trapping at high incidence translational energies. Whereas experimental trapping probabilities at low incidence energies can be correctly predicted once the energy landscape matches the experimental binding energy, the same is not the case at high incidence energies. In agreement with previous literature, we conclude that this failure is likely due to the neglect of strong nonadiabatic effects that arise from transient charge- and/or spin-transfer between metal and molecule yielding a change in effective energy landscape, which goes beyond nonadiabatic energy dissipation that is described via electronic friction theory. To resolve this failure of electronic friction theory, stochastic surface hopping methods such as IESH will likely be required. However, such methods need to be integrated with more realistic first-principles determined charge transfer states[50], which remains very challenging for periodic metallic systems. To fully understand the case of NO on Au(111), additional experiments that provide insight into the trapping probability of highly vibrationally excited molecules at high incidence energies will be useful in the future.

We believe that the here presented approach and simulation results, together with the extensive experimental data by Wodtke and coworkers, provide a firm baseline for the future development of more reliable and efficient nonadiabatic dynamics methods that will open the door to study nonadiabatic effects in thermal and photoelectrochemical reactions at catalyst surfaces in the future.

## COMPUTATIONAL METHODS

In Ref. [40], we have performed spin-polarized DFT calculations for the NO + Au(111) system using the Vienna Ab initio Simulation Package[51, 52] with the PW91 functional.[53] The Au(111) surface was represented by a four-layer slab model in a 3×3 unit cell with the top two layers movable. The Brillouin zone was sampled by a 4×4×1 Gamma-centered $k$-point grid. A total of 2722 points with both energies and forces were collected mainly from direct dynamics trajectories to represent the adiabatic PES. More details can be found in Ref. [40]. In this work, additionally, the 6×6 ODF electronic friction tensor (EFT) was evaluated for 1647 (+ 1052) training (+ test) points using our implementation within the all-electron numerical atomic orbital code FHI-Aims.[12, 54] In the SI, we provide further details on numerical settings and evidence of the robustness of the friction tensor evaluation with respect to these settings (see Figure S3 in the Supplementary Information (SI)). Quasi-classical trajectory calculations were performed using a modified VENUS code.[55]

We employ the recently developed embedded atom neural network (EANN) approach to represent the scalar potential energy[56] and EFT[41] surfaces for NO on Au(111). In the EANN model, the potential energy is expressed as the sum of the embedded atomic energy, each of which is a complex function of the embedded density of the corresponding central atom. Different from potential energy, the EFT is covariant with respect to rotation (or reflection) of the molecule and permutation of identical atoms in the molecule, which is much more difficult to learn by neural networks. For the NO + Au(111) system, we start with a 6 × 3 first-order derivative matrix (corresponds to three neurons in the NN output layer) and a 6 × 6 second-order derivative matrix in terms of the partial derivatives of neural network outputs with respect to atomic Cartesian coordinates of the NO molecule. Multiplying the first- and second-order derivative matrices with their own transpose respectively yields two 6 × 6 matrices that naturally guarantee the rotational covariance and positive semidefiniteness of the EFT. The summation of the two 6 × 6 matrices is employed to approximate the EFT to account for additional symmetry of the EFT with respect to a symmetric mirror. More technical details can be found in our recent work[41, 56-58] and in the SI.

## ASSOCIATED CONTENT

### Supporting Information

The Supporting Information is available free of charge on the ACS Publications website. It includes further computational method details, convergence data and additional results (PDF).

All orbital-dependent electronic friction calculations for training and testing are available within the NOMAD repository (https://dx.doi.org/10.17172/NOMAD/2020.10.06-1). For ease of comparison, the model data points for all final vibrational state and trapping figures are available on figshare (https://dx.doi.org/10.6084/m9.figshare.13049687).

## AUTHOR INFORMATION

**Corresponding Author**

\***Reinhard J. Maurer** – Department of Chemistry, University of Warwick, Gibbet Hill Road, CV4 7AL, United Kingdom; Email: r.maurer@warwick.ac.uk

\***Bin Jiang** - Hefei National Laboratory for Physical Science at the Microscale, Department of Chemical Physics, Key Laboratory of Surface and Interface Chemistry and Energy Catalysis of Anhui Higher Education Institutes, University of Science and Technology of China, Hefei, Anhui 230026, China; Email: bjiangch@ustc.edu.ch

**Author**

**Connor L. Box** - Department of Chemistry, University of Warwick, Gibbet Hill Road, CV4 7AL, United Kingdom; https://orcid.org/0000-0001-7575-7161

**Yaolong Zhang** - Hefei National Laboratory for Physical Science at the Microscale, Department of Chemical Physics, Key Laboratory of Surface and Interface Chemistry and Energy Catalysis of Anhui Higher Education Institutes, University of Science and Technology of China, Hefei, Anhui 230026, China

**Rongrong Yin** - Hefei National Laboratory for Physical Science at the Microscale, Department of Chemical Physics, Key Laboratory of Surface and Interface Chemistry and Energy Catalysis of Anhui Higher Education Institutes, University of Science and Technology of China, Hefei, Anhui 230026, China

**Author Contributions**

The manuscript was written through contributions of all authors. All authors have given approval to the final version of the manuscript. ‡These authors contributed equally.

## ACKNOWLEDGMENT

C. B. is supported with an EPSRC-funded Ph.D. studentship. R.J.M acknowledges funding from the UKRI Future Leaders Fellowship program (MR/S016023/1). B. J. is supported by National Key R&D Program of China (2017YFA0303500) National Natural Science Foundation of China (21722306, 91645202, 22073089), Anhui Initiative in Quantum Information Technologies (AHY090200). C. B. and R. J. M. acknowledge high performance computing resources granted by the Scientific Computing Research Technology Platform of the University of Warwick and access to ARCHER via the EPSRC-funded Materials Chemistry Consortium (EP/R029431), while B. J. appreciates these resources in the Supercomputing Centre of University of Science and Technology of China.

# Supporting Information:

# Determining the effect of hot electron dissipation on molecular scattering experiments at metal surfaces

Connor L. Box[1#], Yaolong Zhang[2#], Rongrong Yin[2], Bin Jiang[2,*], and Reinhard J. Maurer[1,*]

[1]*Department of Chemistry, University of Warwick, Gibbet Hill Road, Coventry, CV4 7AL, United Kingdom*

[2]*Hefei National Laboratory for Physical Science at the Microscale, Department of Chemical Physics, Key Laboratory of Surface and Interface Chemistry and Energy Catalysis of Anhui Higher Education Institutes, University of Science and Technology of China, Hefei, Anhui 230026, China*

[#]: These authors contributed to this work equally.

[*]: corresponding authors: R.Maurer@warwick.ac.uk; bjiangch@ustc.edu.cn

Table of Contents

### Neural network representations
**Potential energy surface**

To enable highly efficient molecular dynamics simulations with electronic friction (MDEF), we developed analytical representations for the adiabatic potential energy surface (PES), the local density friction approximation (LDFA) based scalar friction and orbital-dependent friction (ODF) based electronic friction tensor (EFT), by means of the recently developed embedded atom neural network (EANN) approach. This machine learning approach has been successfully applied to fit scalar and tensorial properties of molecular and periodic systems with high accuracy and efficiency.[1-5] Like other popular atomistic machine learning methods for atomistic simulations,[6] in the EANN model, the total energy of the system is written as the sum of atomic energies, each of which can be expressed as a function of the embedded density of corresponding center atom,

$$E = \sum_{i=1}^{N} E_i = \sum_{i=1}^{N} NN_i(\boldsymbol{\rho}^i). \tag{S1}$$

Here we use atomic neural networks to represent the relationship between embedded density and embedded energy and the embedded density-like descriptors $\boldsymbol{\rho}^i$ can be evaluated by the square of linear combination of simple Gaussian-type atomic orbitals of neighbor atoms.

$$\rho_{L,\alpha,r_s}^{i} = \sum_{l_x,l_y,l_z}^{l_x+l_y+l_z=L} \frac{L!}{l_x!l_y!l_z!} (\sum_{j=1}^{n_{\text{atom}}} c_j \varphi_{l_xl_yl_z}^{\alpha,r_s}(\mathbf{r}^{ij}))^2, \tag{S2}$$

where $\varphi_{l_xl_yl_z}^{\alpha,r_s}(\mathbf{r}^{ij})$ represent Gaussian-type orbitals,

$$\varphi_{l_xl_yl_z}^{\alpha,r_s}(\mathbf{r}^{ij}) = x^{l_x} y^{l_y} z^{l_z} \exp(-\alpha|r-r_s|^2) \tag{S3}$$

where $\mathbf{r}=(x, y, z)$ is the coordinate vector of an electron with its origin at the nucleus of a neighboring atom, $r$ is the norm of the vector, $\alpha$ and $r_s$ are parameters that govern radial distributions of atomic orbitals, $l_x+l_y+l_z=L$ specifies the orbital angular momentum ($L$). Since these density-like descriptors are evaluated by the linear summation of atomic orbitals, the computational cost of the EANN model scales linearly with the number of neighboring atoms.

The DFT energy and force data for representing the PES were taken from our previous work for the NO + Au(111) system[7] and will not be discussed here. The final EANN PES uses 33 descriptors for N/O and 33 for Au, with two hidden layer (40×50) NNs for each type of atom. It should be noted that, the PES in Ref. [7] was trained by the Behler-Parrinello type of atomistic neural network (AtNN) method.[8] We choose the EANN approach to refit the PES because of its high efficiency. We find that the EANN PES is about 2.8 times faster than the previous AtNN PES. Concerning the accuracy, the overall root mean square errors (RMSEs) between the DFT data and the EANN PES are 1.2 meV/atom and 36.2 meV/Å for energy and forces, respectively, comparable to those for the AtNN PES with errors 0.9 meV/atom and 31.6 meV/Å. In addition, the stationary points along the minimum energy path are quite similar on the two PESs, as shown in Figure S1a. Figure S1b further compares the final vibrational state distributions of NO($v_i$=16) calculated by the EANN and AtNN PESs, respectively, which almost coincide with each other. These results indicate that the new EANN PES should retain all the dynamics discovered in Ref. [7] and can serve as the starting point to discuss the nonadiabatic effects introduced by electronic friction.

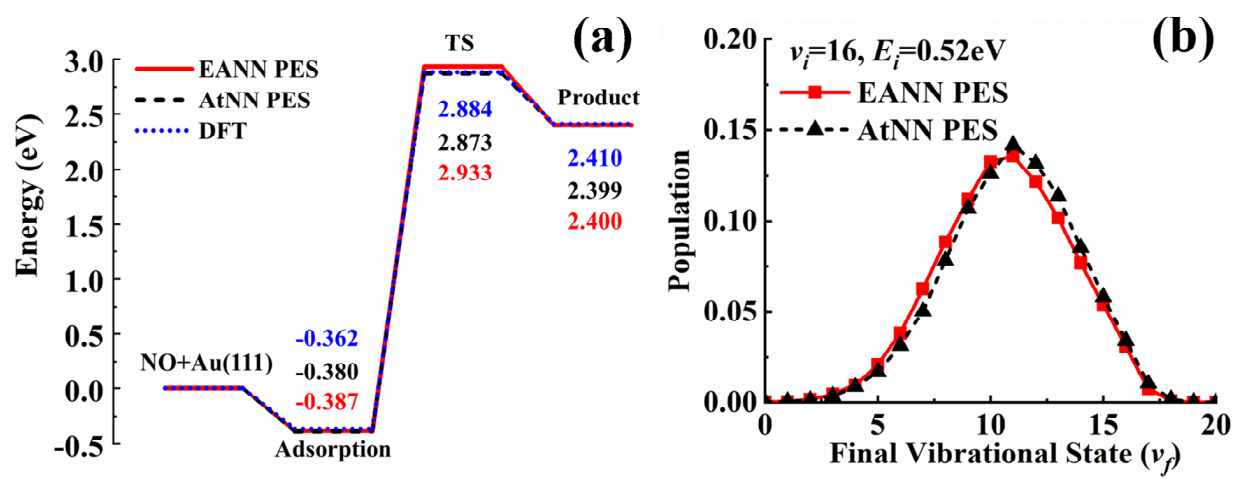

Figure S1 (a) Energetics of the stationary points along the minimum energy path calculated by EANN PES, AtNN PES and DFT. (b) Comparison of vibrational state distributions of NO($v_i$=16) scattering from Au(111) calculated based on EANN and AtNN PESs at $E_i$=0.52 eV.

**Electronic friction tensor**

The LDFA based friction coefficients rely on the bare surface density at each atomic position only.[9] In this regard, the electron density surface (EDS) as a function of surface atom displacements is formally equivalent to the PES of a single atom adsorbed on a surface.[7] Following our previous work,[7] the constructed EANN EDS model contains 18 descriptors for N/O/Au with two hidden layer (20×30) NNs for each type of atom, which well reproduced the DFT charge densities with an extremely low RMSE of $3.05\times10^{-4}$ e/Å$^{-3}$. This level of accuracy is more than sufficient for our purpose.

Differing from the scalar friction coefficient as a result of LDFA, the ODF based full-rank EFT is covariant with respect to rotation (or reflection) of the molecule and permutation of identical atoms in the molecule, which needs to be fulfilled in the neural network representation. We recently proposed a tensorial EANN model to preserve the correct symmetry and the covariant property of EFT inspired from its perturbation theory expression,[2] namely the square of the first derivative nonadiabatic coupling terms among different Kohn-Sham (KS) orbitals.[10] For a molecule with $N$ atoms interacting with a surface, we first crafted a $3N \times M$ first-order derivative matrix ($\mathbf{D}^{NN_1}$) in terms of the partial derivatives of the EANN outputs with respect to $3N$ atomic Cartesian coordinates of the molecule ($\mathbf{R}$). Specifically, $\mathbf{D}^{NN_1}$ can be written in the following form,

$$\mathbf{D}^{NN_1} = \sum_{i=1}^{N} \mathbf{D}^{NN_1,i} = \sum_{i=1}^{N} \frac{\partial \mathbf{H}}{\partial \boldsymbol{\rho}^i} \frac{\partial \boldsymbol{\rho}^i}{\partial \mathbf{R}}. \tag{S4}$$

where $\mathbf{H}$ is an atomic NN output vector with $M$ components ($M$ should be no less than $3N$). But this term is always symmetric under reflection of the molecule about any symmetry plane. In order to mimic the anti-symmetry of KS orbitals, we introduced another $3N \times 3N$ second-order derivative matrix ($\mathbf{D}^{NN_2}$), in terms of second derivatives of density descriptors ($\boldsymbol{\rho}^i$) with respect to $\mathbf{R}$ multiplied with the first derivatives of atomic NN outputs with respect to descriptors, namely

$$\mathbf{D}^{NN_2} = \sum_{i}^{N} \mathbf{D}^{NN_2,i} = \sum_{i}^{N} \sum_{l=1}^{n_{orb}} \mathbf{S}^{il} F^{il}, \tag{S5}$$

where $n_{orb}$ is the number of density descriptors of atom $i$, $\mathbf{S}^{il}$ is the second derivative matrix of the $l$th density descriptor of atom $i$ ($\rho^{il}$) with respect to $\mathbf{R}$,

$$\mathbf{S}^{il} = \nabla_{\mathbf{R}}^2 \rho^{il}, \tag{S6}$$

and $F^{il}$ is first derivative of the EANN output with respect to $\rho^{il}$,

$$F^{il} = \frac{\partial H}{\partial \rho^{il}}. \tag{S7}$$

Multiplying $\mathbf{D}^{NN_1}$ and $\mathbf{D}^{NN_2}$ with their own transpose matrix yields two $3N \times 3N$ matrices that naturally guarantee the rotational covariance and positive semidefiniteness of the EFT,

$$\mathbf{\Lambda}^{NN_i} = (\mathbf{D}^{NN_i})^T \mathbf{D}^{NN_i} \quad (i=1,2). \tag{S8}$$

Our final expression of the NN-based EFT is the sum of the two $3N \times 3N$ matrices $\mathbf{\Lambda}^{NN_1}$ and $\mathbf{\Lambda}^{NN_2}$ which mixes the symmetric and anti-symmetric properties of KS orbitals with respect to a certain symmetry plane, respectively. More technical details can be found in Ref. [2].

The computational details of ODF based EFT for the NO + Au(111) system have been given in the main text. To cover the configuration space, we calculated EFTs for 1647 points extracted from the training set for the PES fitting. These points were used to train the EANN EFT model in which 52 descriptors and three NNs with $40 \times 40 \times 6$ structures were employed. An additional 1052 configurations randomly sampled from multiple classical trajectories served as test set. Figure S2a displays correlation diagrams of DFT calculated ODF and the EANN EFTs for these 1052 configurations not included in the training set. As can be seen, the test RMSE (~0.029 ps$^{-1}$) is very close to the training RMSE (~0.028 ps$^{-1}$), which is small enough relative to the numerical range of EFT elements. In Figure S2b, we monitor the variation of the EANN EFT along a representative trajectory initiated from NO($v_i$=3). The diagonal friction tensor element for internal stretch ($\Lambda_{rr}$) and the off-diagonal coupling between internal stretch and translation ($\Lambda_{rZ}$) along the trajectory were computed by DFT and the EANN model. The excellent agreement further validates the generalizability of the EANN EFT model.

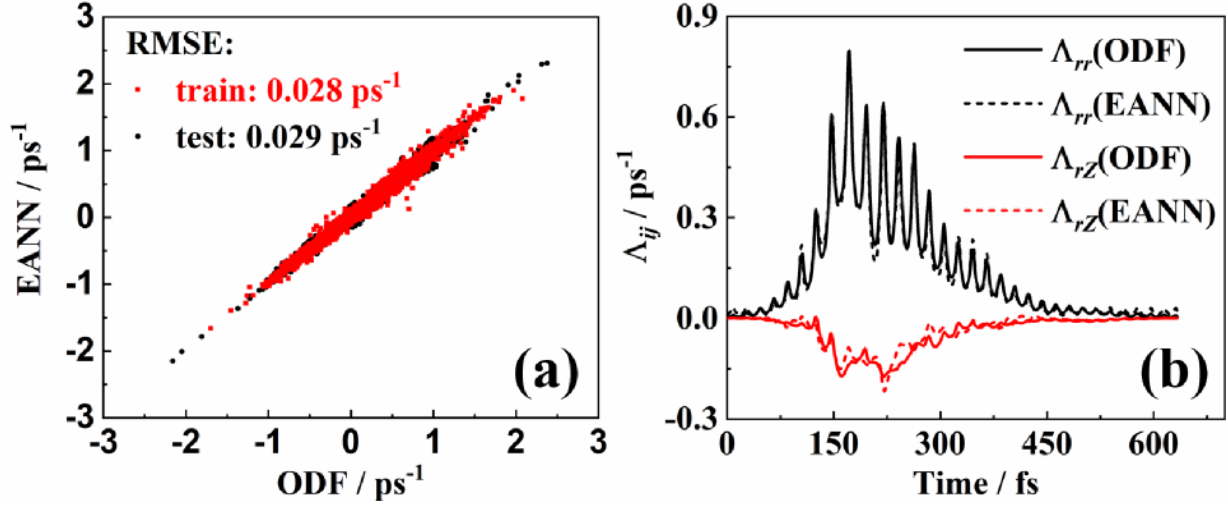

Figure S2. (a) Correlation plots of EFT between the DFT-calculated ODFT and the EANN EFT. (b) Comparison of EFT elements $\Lambda_{rr}$ and $\Lambda_{rZ}$ calculated by ODF and EANN along one representative trajectory for NO($v_i$=3) scattering on Au(111).

**Rescaling PES to experimental binding energy**

In order to check the influence of the overestimated adsorption well on the trapping probability, we rescaled the adsorption energy of the PES to the experimental binding energy (0.24 eV)[11] by adding a switching function $E_{corr}$ (in eV),

$$E_{corr}(Z) = \begin{cases} 0.15, & Z \leq 2.75 \\ 0.075 \times (\cos((Z-2.75)/2.25 \times \pi)+1), & 2.75 < Z \leq 5 \\ 0, & Z > 5 \end{cases} \quad (S9)$$

where $Z$ (in Å) is the center of mass of NO in the Z direction (see Figure 1).

**ODF EFT scaling method**

Presented in the main manuscript (Figure 5b-c) are final state distributions predicted by ODF with the internal stretch element scaled by 4 and LDFA with all coefficients scaled by 4. Additionally shown in Figure S9 and S10 are results with the whole ODF EFT scaled. Scaling the whole EFT or LDFA coefficients is trivial, but selectively scaling the internal stretch element requires a transformation of the EFT from Cartesian to internal coordinates. The transformation matrix ($U$) is given by:

$$U = \begin{bmatrix} \frac{(x_1-x_2)m_{r2}}{r} & \frac{(x_1-x_2)(z_1-z_2)m_{r2}}{r_1} & -(y_1-y_2)m_{r2} & 1 & 0 & 0 \\ \frac{(y_1-y_2)m_{r2}}{r} & \frac{(y_1-y_2)(z_1-z_2)m_{r2}}{r_1} & (x_1-x_2)m_{r2} & 0 & 1 & 0 \\ \frac{(z_1-z_2)m_{r2}}{r} & -r_1 m_{r2} & 0 & 0 & 0 & 1 \\ \frac{(x_2-x_1)m_{r1}}{r} & \frac{(x_2-x_1)(z_1-z_2)m_{r1}}{r_1} & -(y_2-y_1)m_{r1} & 1 & 0 & 0 \\ \frac{(y_2-y_1)m_{r1}}{r} & \frac{(y_2-y_1)(z_1-z_2)m_{r1}}{r_1} & (x_2-x_1)m_{r1} & 0 & 1 & 0 \\ \frac{(z_2-z_1)m_{r1}}{r} & r_1 m_{r1} & 0 & 0 & 0 & 1 \end{bmatrix} \quad (S10)$$

Where:

$m_1$ and $m_2$ are the mass of O and N respectively, $x_1$, $y_1$, $z_1$ ($x_2$, $y_2$, $z_2$) are the Cartesian coordinates of O (N) and

$$m_t = m_1 + m_2$$

$$m_{r1} = \frac{m_1}{m_t}$$

$$m_{r2} = \frac{m_2}{m_t}$$

$$r = \sqrt{(x_1-x_2)^2 + (y_1-y_2)^2 + (z_1-z_2)^2}$$

$$r_1 = \sqrt{(x_1-x_2)^2 + (y_1-y_2)^2}$$

To transform the cartesian EFT ($\Lambda^C$) to internal coordinates ($\Lambda^{IC}$):

$$\Lambda^{IC} = U^T \Lambda^C U$$

Once in internal coordinates, the internal stretch element ($\Lambda_{rr}^{IC}$) can be selectively scaled by 4. The transformation of the internal coordinate EFT and back to Cartesian is given by:

$$\boldsymbol{\Lambda}^C = \boldsymbol{U}^{T^{-1}} \boldsymbol{\Lambda}^{IC} \boldsymbol{U}^{-1}$$

### ODF EFT training data
**Calculation details**

The same slab model is used as the underlying adiabatic PES training data, while an extended vacuum space (70 Å) in *z* direction was imposed to remove the slight interaction of vertically repeated slabs. The current implementation employs a finite difference approach (a Cartesian displacement of 0.0025 Å is used) to evaluate the first order response of Hamiltonian and overlap matrices. At each displacement, a self-consistent DFT calculation is carried out using the PBE exchange correlation functional with tolerances for the total energy, the eigenvalue energies and the electronic density of $1e^{-6}$ eV, $1e^{-3}$ eV and $1e^{-5}$ $e/a_0^3$ respectively. A standard 'tight' numerical basis set is employed with a 9×9×1 *k*-point mesh. The EFT is evaluated using a Gaussian smearing function of width 0.6 eV and a Fermi factor corresponding to an electronic temperature of 300 K. Additionally, an atomic ZORA relativistic correction and a 0.1 eV width Gaussian occupation smearing was employed.

**Convergence of ODF EFT**

The convergence of the ODF EFT was investigated for the 4 above described MERP structures (reactant, adsorption, transition state and product) with respect to basis set size ('light', 'tight' and 'really tight' presets in FHI-Aims), *k*-point mesh size and number of substrate layers (up to an additional 2). Convergence behavior of the internal stretch EFT element with respect to *k*-point mesh size is shown in Figure S3a, where we conclude a 9×9×1 grid is reasonably converged and computationally efficient. The inclusion of a dipole correction was tested and had no significant effect on the EFT values. The EFT for the employed lattice constant as well as the PBE optimized lattice constant for a top site geometry did not significantly differ. Additionally, the stability of EFT values was investigated with respect to the width of the Gaussian function employed within this implementation of EF (Figure S3b). Whilst justifications for various broadening widths have been put forward in literature,[10, 12, 13] this added dependency is an undeniable weakness in the implementation but currently necessary to obtain stable sampling with practical **k**-point mesh sizes.

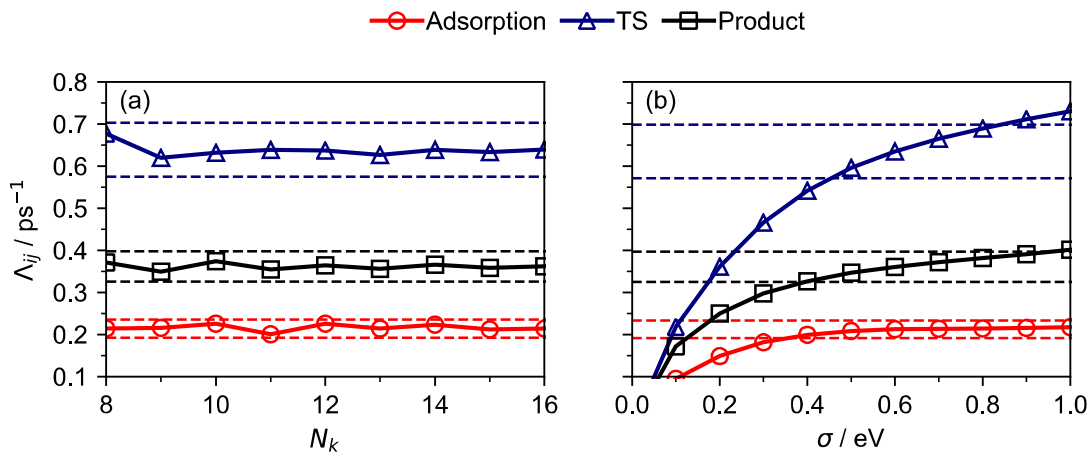

Figure S3. Convergence of the mass-weighted internal stretch friction tensor element for the adsorption, TS and product geometries from the MERP with respect to (a) **k**-point mesh size ($N_k \times N_k \times 1$) and (b) friction gaussian smearing width ($\sigma$). Dashed lines on (a) show +/- 10 % from the final (16×16×1) grid value, whilst on the (b) they show +/- 10 % from the 0.6 eV smearing value.

**Friction spectrum calculation details**

The mass-weighted internal stretch friction excitation spectrum presented in Figure 5a is calculated using the same settings as the ODF training data, except a denser **k**-point mesh of 11×11×1 was used to smoothly represent the spectrum.

**Effect of intraband excitations**

As mentioned in the main manuscript, the couplings for intraband and higher order excitations are excluded. A crude estimation of the effect of including these couplings is to increase the unit cell size to fold these intraband excitations into interband excitations within the first Brillouin zone that are captured by our evaluation of the ODF tensor. The behavior is investigated for p(2×2), p(3×3) (employed in the training data), p(4×4) and p(5×5) unit cells for the 'reactant', 'adsorption' and 'transition state' geometries defined above. The calculation used the same settings as for the training data except a $N_k$=18, 14, 9 and 5 ($N_k \times N_k \times 1$) $\boldsymbol{k}$-point mesh was used respectively for the p(2×2), p(3×3), p(4×4) and p(5×5) unit cells, the last of which required a light basis set.

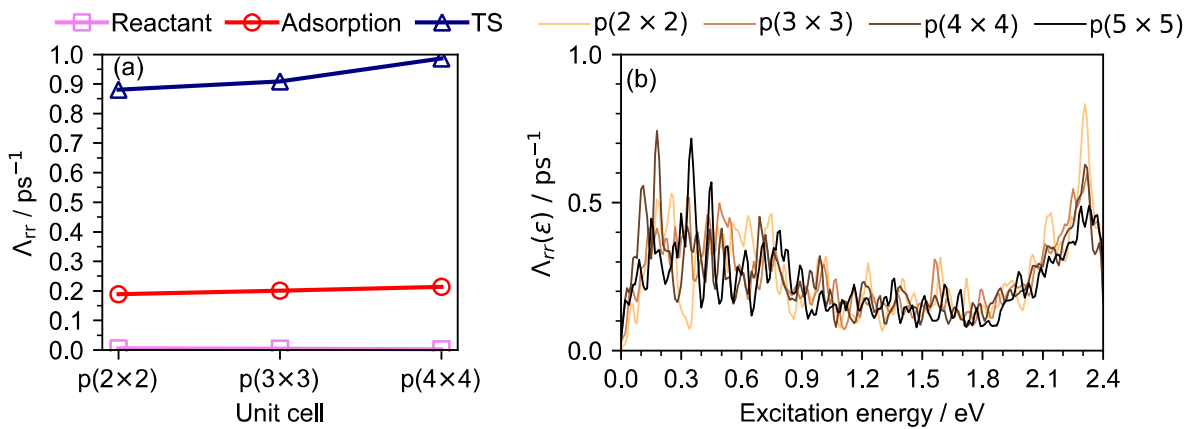

Figure S4. (a) Size of mass-weighted internal stretch friction tensor element for 3 structures (defined above) with respect to unit cell size. (b) Mass-weighted internal stretch friction excitation spectrum for the adsorption structure, plotted with 3 different unit cell sizes.

Figure S4a shows the internal stretch element from the EFT is fairly insensitive to unit cell size. This shows the effect of excluding intraband excitations in this regime is weak and effectively washed out by the 0.6 eV broadening employed. As can be seen in Figure S4b, increasing the unit cell does not significantly alter the high energy regimes but does increase the available low energy excitations, thereby increasing the observed EF. The unit cells employed specifically in this section are not relaxed and employ the experimental lattice constant, thus the p(3×3) spectrum in Figure S4b differs slightly from that presented in Figure 5 in the main manuscript.

**Dynamics details**

Quasi-classical trajectory (QCT) calculations were performed using a modified VENUS code[14]. The NO molecules were initialized from 6.0 Å above the surface with the molecular center of mass in the 3×3 unit cell. The initial vibrational momentum of the NO molecule was selected using the standard normal mode sampling scheme[15]. The NO molecule was then considered as a rigid rotor, whose rotational angular momentum, $\bar{\mathbf{j}} = \sqrt{J(J+1)}\hbar$ with the rotational quantum number $J$ was acted on the molecule, followed by random orientation of NO before impinging to the surface along the surface normal at a given incident energy. The Au(111) surface was first equilibrated to 300 K for 5 ps by the Andersen thermostat[16]. The velocity Verlet algorithm was employed to propagate the trajectories on the surface up to 10 ps with a 0.10 fs time step. A trajectory with the NO molecule sticking on the surface after 10 ps was considered "trapped", otherwise "scattered" when the molecule is reflected back to the vacuum. For scattered molecules, the vibrational action number $v_f$ and rotational quantum number $j_f$ were calculated by Einstein−Brillouin−Keller (EBK) semi-classical quantization and rotational angular momentum respectively[17]. The final quantum numbers were obtained by standard histogram binning (HB), which rounds each action number to its nearest integer. An alternative way of binning is to weight each action number with a narrow Gaussian weighting function centered on the integer quantum number, often referred as the Gaussian binning (GB) scheme.[18, 19]

$$P_{GB}(v) = G(v)/N_{total}^{G}, \qquad (S11)$$

where $N_{total}^{G}$ is the sum of $G(v)$ of all trajectories. The Gaussian weight is defined as,

$$G(v) = \frac{\beta}{\sqrt{\pi}} e^{-\beta^2 (v'-v)^2}, \qquad (S12)$$

We have compared the results of GB ($\delta$ =0.1) and HB in Figure S3 for NO($v_i$=16). Indeed, both binning methods yield very similar vibrational state distributions, while GB requires many more trajectories to achieve good statistics.

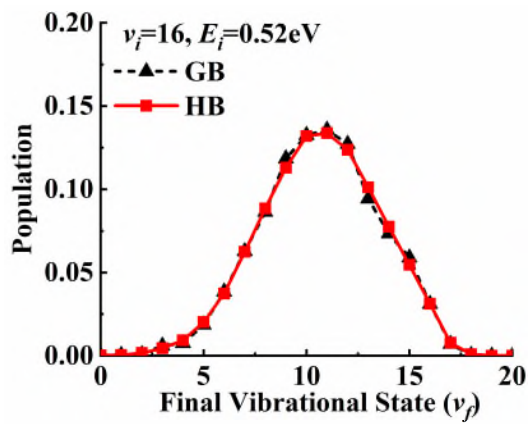

Figure S5. Comparison of calculated vibrational state distributions of NO($v_i$=16) scattering from Au(111) obtained by GB and HB at $E_i$=0.52 eV.

**Orientation details**

To shed light on the failure to capture multi-quantum loss, trajectories with different initial orientations were analyzed. For simplicity, the N-down and O-down configurations were separated given their relative heights of N and O atoms, i.e. N-down for $Z_N < Z_O$ and O-down for $Z_N > Z_O$, which does not exactly correspond to (but is similar to) the experimentally aligned distributions in the electric field[20]. This is sufficient for our purpose and the steric effects will be discussed in more detail in a forthcoming publication with exactly sampled initial orientation distributions.

**Angular distribution**

The experimental angular distribution reported for $v_i$=3, $j_i$=0, $E_i$=0.520 eV suggests a short residence time of the scattered NO on the surface and thus a multibounce dominated scattering mechanism is unlikely. The bounce dynamics for this system have previously been studied by other groups[21, 22] and some of us.[7] In this work, the number of bounces is classified as the number of times the z component of the NO COM crosses an xy plane upwards at 3 Å height from the surface. In Figure S6, we present scattering angle distributions for both single bounce and all bounce trajectories, which are broader than the experimental cosine fit with power of 8-9,[21] but much narrow than the $\cos(\theta)$ curve shown expected from a long residence time.[23] The negligible difference between the distributions for single bounce and all bounce trajectories suggest multibounce scattering plays only a small role for these initial conditions. Figures 2-5 in the main manuscript included only single bounce scattering trajectories, which is discussed in the next section.

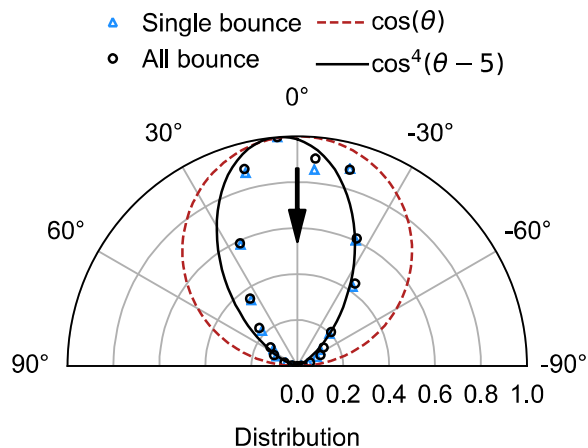

Figure S6. Normalized scattering angle (with respect to surface normal) distributions for $v_i=3$, $j_i=0$, $E_i=0.520$ eV using ODF for both single bounce only trajectories and for all scattered trajectories. Dashed line shows $\cos(\theta)$, whilst the solid line shows a $\cos^4(\theta-5)$ fit to the all bounce data. Incident angle= $0°$ (shown by black arrow) for all trajectories.

**Additional population data**

Final state populations as reported in the main manuscript were renormalized to only include final states that were reported in experiment. The reported final state distributions from state-to-state scattering of Wodtke and coworkers do not include $v_f=0$, $v_f>v_i$ or $v_f=1$ (when $v_i>10$) states for practical reasons and reasons of scope,[24] which is why we exclude these final states in our analysis in Figure 2. Additionally, final state populations in the main manuscript only included single bounce scattering trajectories. For completeness, scattering distributions from Figures 2-5 in the main manuscript are reproduced here in Figures S7-10 with all measured final populations reported and all bounce events included, the distributions remain largely unchanged by this. Additional scaling models are also reported and discussed in the following sections. Figure S9 shows that rescaling the PES does not significantly alter the distribution for any of the isotropic and orientated $v_i=3$ populations. As shown in Figure S10, rescaling the PES does shift the distribution of results slightly further away from experiment for $v_i=11$, but does not significantly alter the distribution for $v_i$=16. Figure S10 shows the state distributions for scaling ODF EFT along the intra-molecular stretch vibration DOF of NO by four (as employed in Figure 5b-c in the main manuscript) yields close to identical results as scaling the entire ODF EFT by 4. This suggests that the internal stretch element of the tensor governs the final vibrational state distribution in the EF nonadiabatic regime.

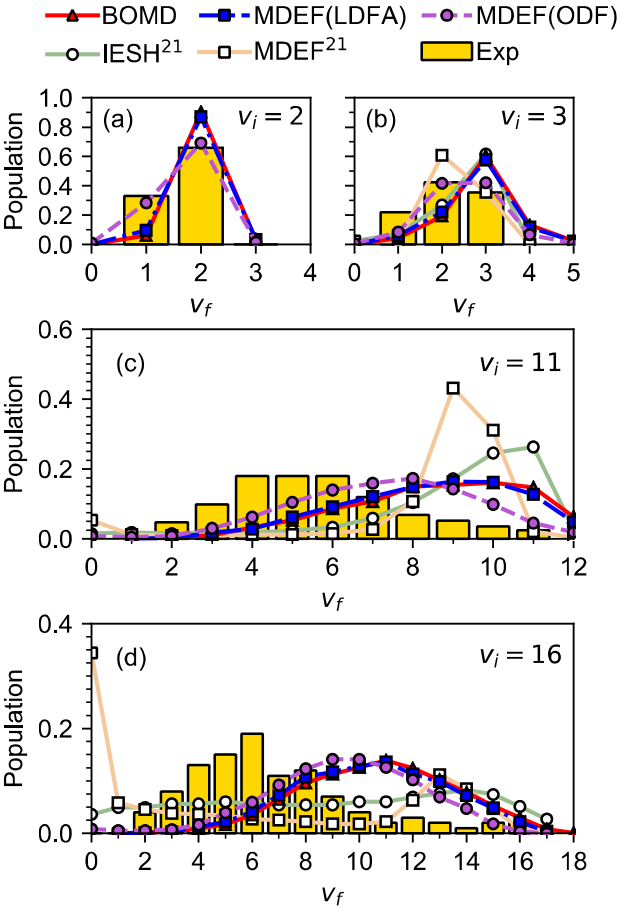

Figure S7. Same as Figure 2 but results from all final vibrational states and all bounce events are included (also applied to reference data[22]).

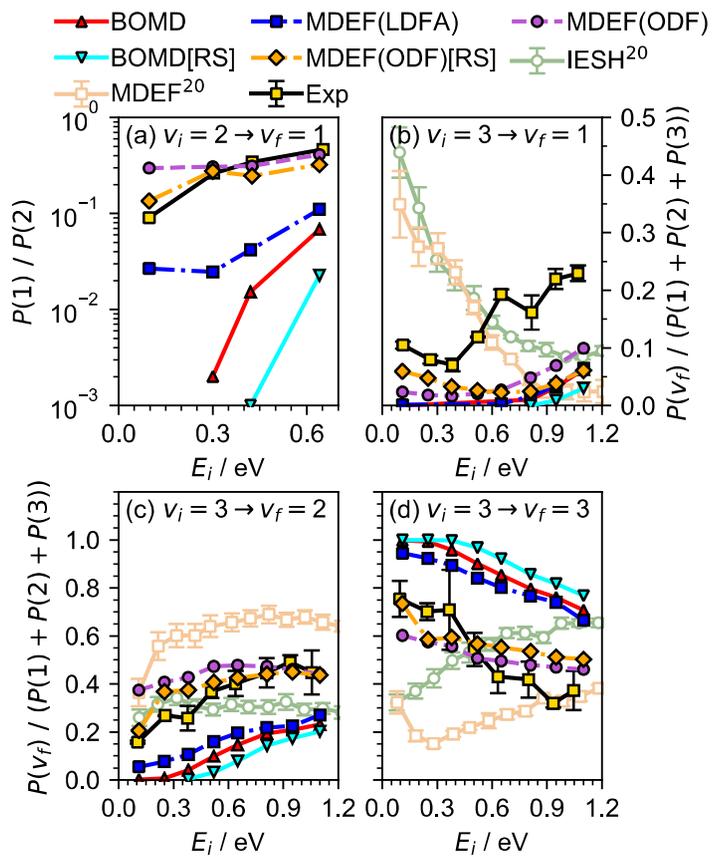

Figure S8. Same as Figure 3 from the main manuscript, except double and multibounce trajectories are not excluded (also applies to reference data[21]) and additional rescaled PES models are shown.

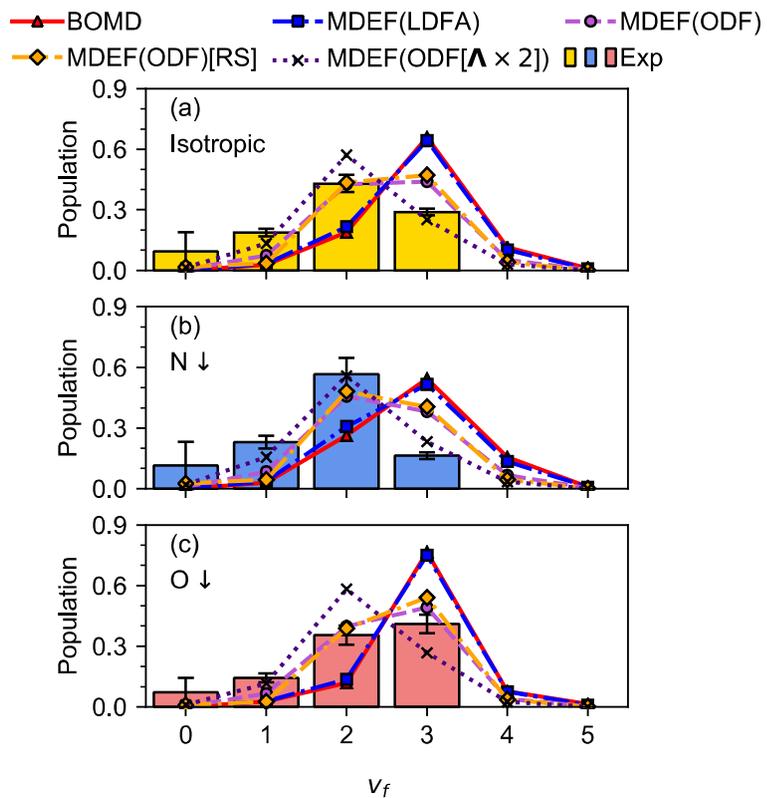

Figure S9. Same as Figure 4b-c but results from all final vibrational states are included and all bounce events are included (also applied to reference data). Also shown are ODF with an adjusted PES and ODF with an isotropic scaling of 2 are included.

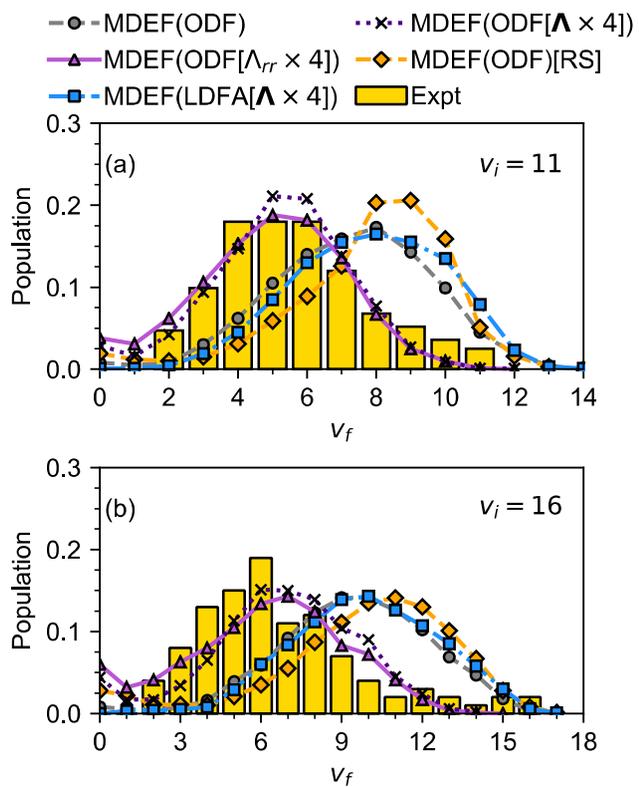

Figure S10. Same as Figure 5b-c but results from all final vibrational states are included and all bounce events. Additionally, shown is the ODF EFT with an isotropic scaling of 4, and ODF with the rescaled PES.

**Analysis of $v_i=3$**
**Dynamical steering effects**

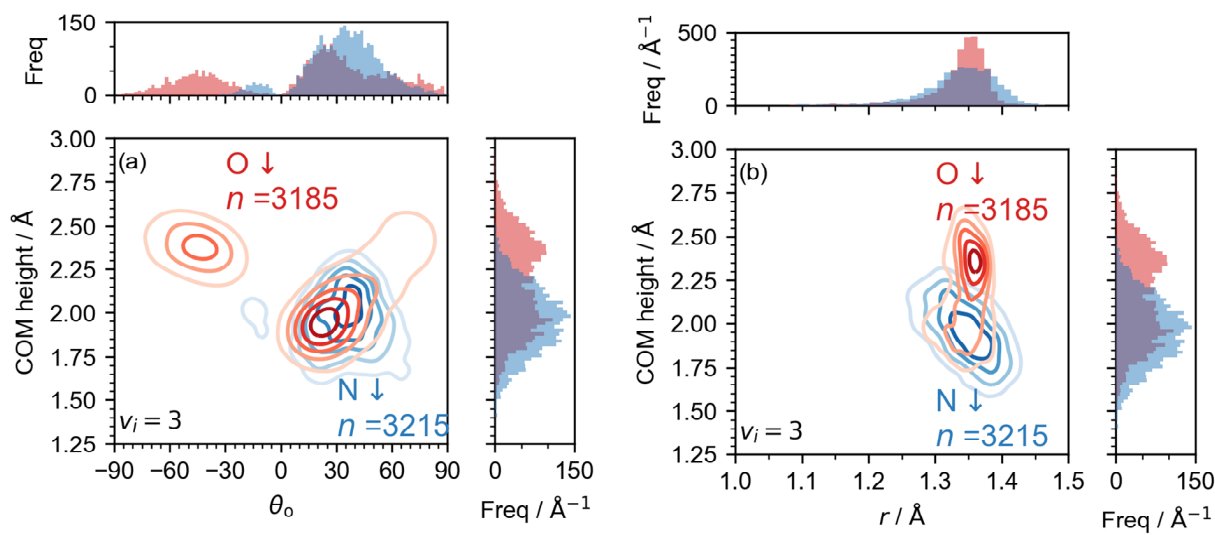

Figure S11. Kernel density estimation plot for (a) surface orientation ($\theta_o$) vs COM height and (b) bond length vs COM height from surface at the point of closest approach for a trajectory within the timeframe of the first bounce with an initial N-down orientation (blue) or O-down (red) for the following incidence parameters: $v_i=3$, $j_i=0$, $E_i=0.950$ eV. Positive angles in (a) correspond to nitrogen orientated towards the surface, whilst negative are the opposite. Plots are annotated with respective total number of trajectories ($n$). Distributions of the trajectories are shown in axes-aligned histogram plots with the same color scheme. ODF model employed. All trajectories included.

Figure S11 shows that even initial O-down orientations preferentially impact with an N-down collision geometry. ($\theta_o > 0$) Only a small amount of initial N-down orientations impact with an O-down collision geometry. ($\theta_o > 0$) On average, initial N-down orientations approach the surface closer.

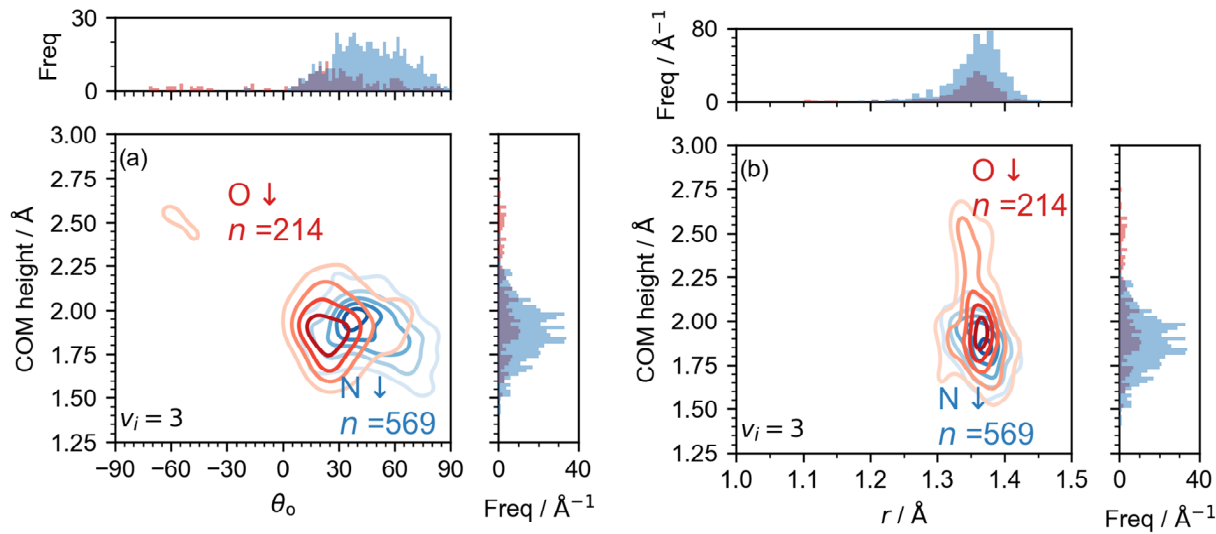

Figure S12. Same as Figure S11 except only trapped trajectories are included.

Figure S12 shows the trapped population (which we expect to be erroneous, see main manuscript) mostly arises from the same region of collision geometries (N-down, and COM height <2 Å, N-O bond lengths of around 1.35 Å). Over twice as many trapped trajectories come from N-down than initial O-down orientations and nearly all trapped trajectories arise from N-down collision geometries.

**Trapping probability**

The trapping population is measured to be negligible at an incidence energy of $E_i$=0.653 eV for $v_i$=2 and $j_i$=2 (see Figure 6 in the main manuscript and experimental reference[25]). As can be seen from Figure S13, trapping predominately occurs on close approach to the surface for the first impact. Figure S13 provides evidence that trajectories with the nitrogen orientated towards the surface ($\theta_o > 0$) and an extended bond length $(1.3 - 1.4 \text{ Å})$ lead NO to the closest possible approach. A high probability of trapping is apparent when the NO is close to the surface (COM ≤ 2.1 Å) and when bond lengths are elongated. Additionally, even when the NO molecule is not initially orientated, the majority of collisions exhibit a steric effect whereby most of the inflection points are N down ($\theta_o > 0$). This reinforces the idea of a significant dynamical steering effect for this system.

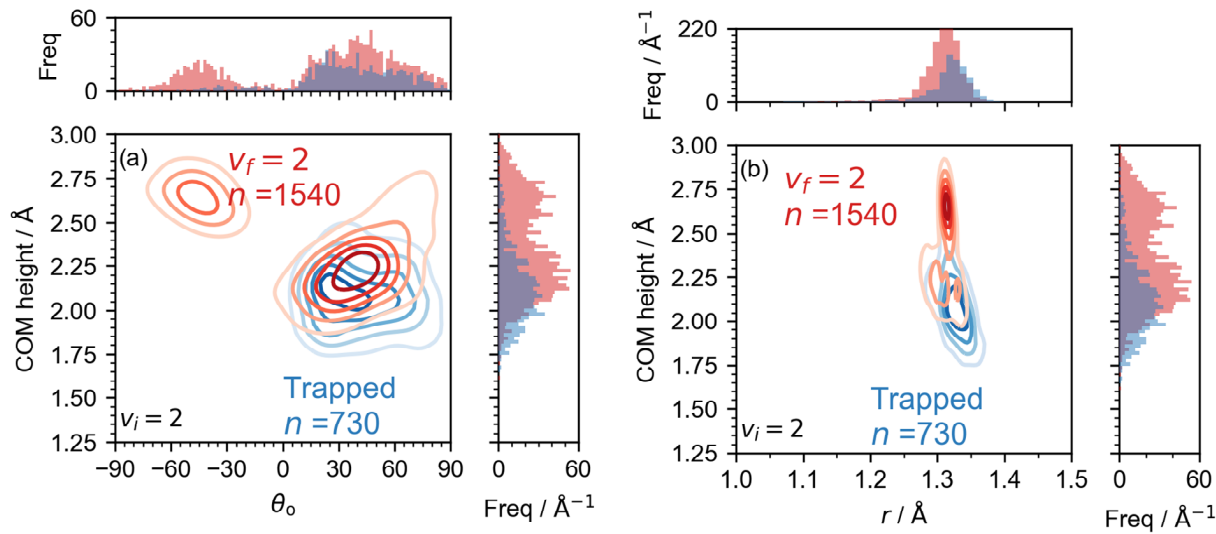

Figure S13. Same as Figure S11 except trapped (blue) or elastic (red) trajectories are considered and the following incidence parameters: $v_i=2$, $j_i=2$, $E_i=0.420$ eV with BOMD model are employed.

**Comparing LDFA and ODF EFT across selected trajectories**

The (position dependent) ODF magnitude is not directly affected by the vibrational energy, however it is indirectly modulated through the positions of the atoms. This is apparent in Figure S14a-b, where the ODF friction tensor value for internal stretch element exhibits peaks during the close approach repulsive part of the vibration (observed by the characteristic fast double peak in the normalized velocities). The ODF internal stretch element reaches magnitudes higher for the higher vibration (shown in Figure S14d) than the lower one (Figure S14b). The LDFA projected internal stretch element, as seen in Figure S14a and c, is far lower in magnitude than ODF, similar findings have been reported in literature.[26] The LDFA value expectedly does not reflect the nature of the vibration at all, but rather just the proximity to the surface. Presented in Figures S14a-b are elastic trajectories, whilst Figures S14c-d undergo multiquanta vibrational energy loss.

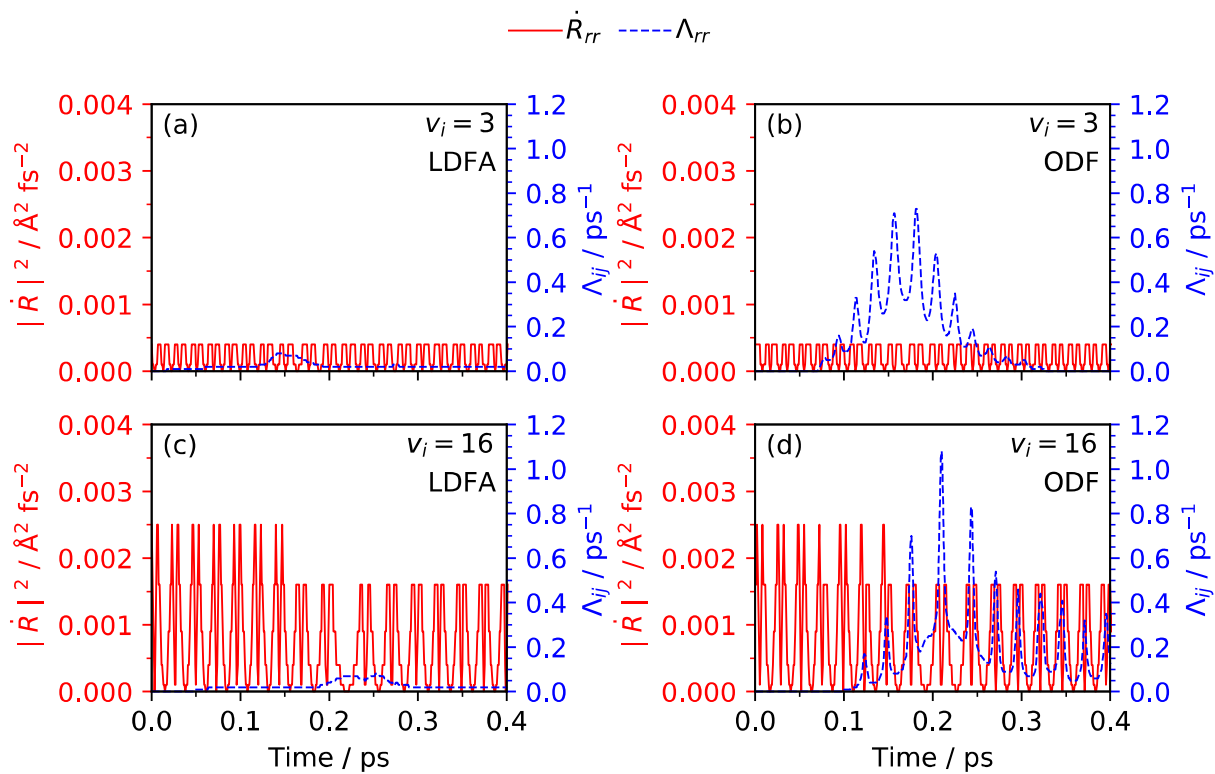

Figure S14. Mass weighted internal stretch projected i) velocity (left scale) and ii) tensor element (right scale) from (a,c) LDFA and (b,d) ODF for (a,b) $v_i=3$, $j_i=0$, $E_i=1.08$ eV and (c,d) $v_i=16$, $j_i=0$, $E_i=0.520$ eV.